\documentclass[conference]{IEEEtran}
\IEEEoverridecommandlockouts
\usepackage{cite}
\usepackage{amsmath,amssymb,amsfonts} 
\usepackage{algorithmic,color,gensymb, soul,xcolor,subcaption}
\usepackage{graphicx}
\usepackage{textcomp}

\def\BibTeX{{\rm B\kern-.05em{\sc i\kern-.025em b}\kern-.08em
    T\kern-.1667em\lower.7ex\hbox{E}\kern-.125emX}}

\begin{document}


\title{Predicting Wireless Channel Features  \\ using Neural Networks}
\author{\IEEEauthorblockN{Shiva Navabi$^{\dagger}$, Chenwei Wang$^*$, Ozgun Y. Bursalioglu$^*$, Haralabos Papadopoulos$^*$}\\
\IEEEauthorblockA{$^{\dagger}$University of Southern California, Los Angeles, CA\\
$^*$DOCOMO Innovations, Inc., Palo Alto, CA\\
 $^{\dagger}$navabiso@usc.edu, $^*$\{cwang, obursalioglu, hpapadopoulos\}@docomoinnovations.com}}
\maketitle

\begin{abstract}
We investigate the viability of using machine-learning techniques for estimating user-channel features at a large-array base station (BS). In the scenario we consider, user-pilot broadcasts are observed and processed by the BS to extract angle-of-arrival (AoA) specific information about propagation-channel features, such as received signal strength and relative path delay. The problem of interest involves using this information to predict the  angle-of-departure (AoD) of the dominant propagation paths in the user channels, i.e., channel features not directly observable at the BS. To accomplish this task, the data collected in the same propagation environment are used to train neural networks. Our studies rely on ray-tracing channel data that have been calibrated against measurements from Shinjuku Square, a famous hotspot in Tokyo, Japan. We demonstrate that the observed features at the BS side are correlated with the angular features at the user side. We train neural networks that exploit different combinations of measured features at the BS to infer the unknown parameters at the users. The evaluation based on standard statistical performance metrics suggests that such data-driven methods have the potential to predict unobserved channel features from observed ones.
\end{abstract}


\section{Introduction}\label{sec:intro}


Cellular standardization efforts have mainly focused on enabling wireless communication in the sub-6GHz spectrum. Given the scarcity of spectrum below 6GHz, new efforts in 3GPP are expanding their scope to include spectrum above 6GHz, in particular the millimeter wave (mmWave) frequencies. To enable the deployment of reliable cost-efficient wireless networks operating at mmWave frequencies, however, a number of serious challenges must be addressed \cite{rangan2014millimeter, ghosh2014millimeter}. This is because compared to its sub-6GHz counterpart, communication at mmWave is impacted by much larger propagation pathloss, more rapidly changing channels, severe penetration loss, dynamic shadowing
, etc. Fortunately, due to shorter wavelengths, much larger arrays can be packed into a small footprint, which enables the use of massive arrays in small-cell BSs and moderately large arrays at the user terminals. Such BS and user-terminal arrays are essential at mmWave as they can be exploited to provide combined transmit-receive (TX-RX) beamforming (BF) gains. Given the harsh propagation conditions at mmWave, such BF gains play a vital role as they can greatly extend the range of mmWave communication, thereby greatly improving coverage.

Learning the user-channels to create beams, however, is more challenging at mmWave. The larger the arrays, the larger the set of beams and TX-RX beam combinations that need to be searched.  As channels decorrelate much faster than at sub-6 GHz bands, high-gain TX-RX beam combinations must be learned much faster. This results in substantially larger training overheads at mmWave due to the need for more frequent searches and within a larger space of beam pairs.
In real-world cellular environments where most user terminals communicate to BSs through channels with no line-of-sight path, having access to precise geometric information regarding the relative locations of the BS and TX arrays (and their relative orientation) is not sufficient to identify TX-RX beam pairs that yield high BF gains.


There is a large body of works on finding the optimal TX-RX beam pair. In \cite{abari2016millimeter}, Abari \emph{et al.} proposed ``Agile-Link" to find the optimal beam alignment through employment of carefully concocted hash functions that can quickly identify and remove the direction bins with no energy. In \cite{hur2013millimeter}, Hur \emph{et al.} proposed an adaptive beam alignment technique where a hierarchical BF codebook is leveraged in lieu of the exhaustive search, but the required feedback from the receiver to the transmitter incurs additional overhead and latency. An attractive  alternative is to estimate the multipath channel parameters \cite{alkhateeb2014channel, schniter2014channel, kim2015virtual, marzi2016compressive, gao2015spatially, zhou2016channel}. Due to the sparse structure of mmWave scattering, \cite{schniter2014channel, kim2015virtual, marzi2016compressive, gao2015spatially, alkhateeby2015compressed} formulated channel estimation as a sparse signal recovery problem, significantly reducing the training overhead in the process. In addition, low-rank tensor factorization has been exploited in \cite{zhou2016channel, zhou2017low} to further improve estimation accuracy and to reduce the computational cost. Based on several recent empirical studies, Li, {\em et al.} showed the existence of a joint sparse and low-rank structure in mmWave channels in presence of angular spreads \cite{li2017millimeter}, and they leveraged this structure to further reduce the computational cost.


While the aforementioned prior works targeted the channel-parameter prediction problem from a system modeling/signal processing perspective, in this paper we leverage the use of machine learning towards this task, so as to improve the user experience and/or the radio-resource utilization in wireless networks. In particular, we exploit machine learning  at the BS to predict user-channel features that are not observable at the BS. Specifically, we consider using AoA dependent propagation-channel features  at the BS (extracted using a  large array) to predict the AoD of the dominant propagation paths in the user channels. The dominant-path AoD prediction is performed by neural networks, which are trained by using the data collected from the same propagation environment. The data of propagation channels that we use were generated by a channel tracer (and also calibrated against measurements) over Shinjuku square, a typical hotspot in Tokyo, Japan.

We recast channel-parameter prediction as a learning-based optimization problem by first constructing appropriate feature representations. Then we conduct correlation analysis between the observed and the unknown features in the model. Next, we train several neural networks, each using different combinations of observed features to infer the unknown parameters at the user side. Our preliminary results suggest that machine learning could  prove a valuable tool in allowing big data to be used for predicting unobserved  channel parameters, and improving network resource utilization and user experience.

\section{System Model and Problem Formulation}\label{sec:sys}

In this section we describe how the channel features that are provided by the ray-tracing data, are turned into observable  and unobservable channel features at the BS side. In the process, we will describe the channel models giving rise to these representations.

We assume half-wavelength spaced uniform linear arrays (ULAs) at the BS and the mobile station (MS) or the user, equipped with $N_{\text{BS}}$ and $N_{\text{MS}}$ antenna elements, respectively. As introduced in \cite{compressed}, a multipath wireless channel can be modeled as a linear system with the following $N_{\textrm{BS}}\times N_{\textrm{MS}}$ time-frequency response matrix:
\begin{eqnarray}\label{eqn:channel_tf}
\mathbf{H}(t,f) = \sum\limits_{l=1}^L \alpha_l\mathbf{a}_{\text{BS}}(\theta_l) \mathbf{a}_{\text{MS}}^H(\phi_l) e^{j 2\pi f\tau_l},
\end{eqnarray}
where $L$ is the number of multipaths, and for the $l^{th}$ path, $\alpha_l$ denotes the complex channel coefficient, $\theta_l \in [0, 2\pi)$ and $\phi_l \in [0, 2\pi)$ represent the associated AoA and AoD, respectively, $\tau_l$ is path delay, and $\mathbf{a}_{\text{BS}}$ and $\mathbf{a}_{\text{MS}}$ are the associated array steering vectors at the BS and the MS, respectively, which can be written as

\small\vspace{-0.15in}
\begin{eqnarray}
\mathbf{a}_{\text{BS}}(\theta_l) &\!\!\!\!=\!\!\!\!& \frac{1}{\sqrt{N_{\text{BS}}}}
\Big[ 1, e^{j \frac{2\pi}{\lambda} d \sin(\theta_l)}, \cdots, e^{j (N_{\text{BS}} - 1) \frac{2\pi}{\lambda} d \sin(\theta_l)}  \Big]^T, \\
\mathbf{a}_{\text{MS}}(\phi_l) &\!\!\!\!=\!\!\!\!& \frac{1}{\sqrt{N_{\text{MS}}}}
\Big[ 1, e^{j \frac{2\pi}{\lambda} d \sin(\phi_l)}, \cdots, e^{j (N_{\text{MS}} - 1) \frac{2\pi}{\lambda} d \sin(\phi_l)}  \Big]^T\!\!,\ \ \ \label{eq:a}
\end{eqnarray}
\normalsize
where $\lambda$ is the carrier wavelength. Assuming the channel follows block fading and considering the channel in a block within coherence time, we can drop the time index $t$. In this paper, for simplicity, we only consider the 2-D system where AoA and AoD are captured by the azimuth components only\footnote{Analysis for the 3-D system can be easily generalized by incorporating elevation components, which would yield channels that are even sparser.}.

According to (\ref{eqn:channel_tf}), $\mathbf{H}(f)$ is determined by the parameters $\{\alpha_l, \tau_l, \theta_{l}, \phi_{l}\}_{l=1}^{L}$, where each is continuous-valued in the corresponding field. However, the observation precision of these parameters is subject to the observation resolution. Specifically, the observation resolution of $\theta_{l}$ and $\phi_{l}$ is limited by the size of the antenna arrays at the BS and the user, respectively, and that of $\tau_{l}$ is determined by the system bandwidth\footnote{Without specifying the bandwidth, we consider the observed delay at the BS can take continuous values by assuming the resolution is infinitely large.}. Hence, after sampling over AoA/AoD domains, the virtual representation of the channel in (\ref{eqn:channel_tf}) is given by \cite{compressed}:

\small\vspace{-0.1in}
\begin{eqnarray}\label{eq:H_dis}
{\bf H}(f)=\sum_{l=1}^L{\bf W}_{\text{BS}}{\bf H}_v^T(l) {\bf W}_{\text{MS}}^T e^{-j2\pi f\tau_l},
\end{eqnarray}
\normalsize
where ${\bf H}_v(l)$ is the $N_{\textrm{BS}}\times N_{\textrm{MS}}$ matrix associated with the $n_r^{\rm th}$ revolvable AoA and $n_t^{\rm th}$ revolvable AoD of the $l^{\rm th}$ path (see equation (4) in \cite{compressed}); ${\bf W}_{\text{BS}}$ and ${\bf W}_{\text{MS}}$ are $N_{\textrm{BS}}\times N_{\textrm{BS}}$ and $N_{\textrm{MS}}\times N_{\textrm{MS}}$ unitary matrices, respectively, which comprise ${\bf a}_{\textrm{BS}}(n_r/N_{\textrm{BS}})$ and ${\bf a}_{\textrm{MS}}(n_t/N_{\textrm{MS}})$ as their $n_r^{\rm th}$ and $n_t^{\rm th}$ columns, respectively. Both ${\bf W}_{\text{BS}}$ and ${\bf W}_{\text{MS}}$ turn out to be DFT matrices with $N_{\textrm{BS}}$, $N_{\textrm{MS}}$ columns, respectively.  Indeed as was shown in \cite{Ansuman-JSDM} for large ULAs projecting the DFT matrix acts like a Karhunen-Loeve expansion, as projecting onto the DFT matrix effectively whitens and sparsifies the channel. Thus, if the BS and the user both apply their respective ${\bf W}_{\text{BS}}$ and ${\bf W}_{\text{MS}}$ as the BF matrices, then we can obtain the effective channel:

\small\vspace{-0.15in}
\begin{align}\label{eq:Heff}
\mathbf{H}_{\text{eff}}(f) = \mathbf{W}^H_{\text{BS}} \mathbf{H}(f) \mathbf{W}_{\text{MS}}=\sum_{l=1}^L{\bf H}_v^T(l)e^{-j2\pi f\tau_l}.
\end{align}
\normalsize
Inspection of (\ref{eq:Heff})  reveals that selecting the optimal beams, i.e., the DFT columns that yield largest-power projections into the AoA/AoD angular bins, is equivalent to seeking the entries of $\mathbf{H}_{\text{eff}}(f)$ with the highest power.

\emph{Notations}: In this paper, we use $|\cdot|$, $\|\cdot \|$ and $\|\cdot \|_{\textrm{F}}$ to denote the amplitude of a scalar, the $\ell_2$-norm of a vector, and the Frobenius norm of a matrix, respectively. Also, given a set $\mathcal{A}$, we use $|\mathcal{A}|$ to denote its cardinality.

\subsection{Problem Formulation}\label{sec:problem}

We refer to the  received signal strength (RSS), multipath delays, (azimuth) AoA bins and AoD bins, corresponding to $r_l=|\alpha_l|^2$, $\tau_l$, $\theta_l$'s and $\phi_l$'s respectively, as features. Note that the BS can observe parameters $\{r_l, \tau_l, \theta_{l}\}_{l=1}^L$ subject to its observation resolution but not their $\phi_{l}$'s since these can only be observed at the user side. Thus, an interesting question arises: \emph{Can the BS extract  features that are only available at the user side, based on what the BS observes?}

While the answer to the question might not be straightforward, let us rethink how the parameters $\{r_l, \tau_l, \theta_{l}, \phi_{l}\}_{l=1}^L$ are generated. As introduced in Sec. \ref{sec:intro}, given a specific area, they are obtained via measuring the signals that travel through the environment where the locations of buildings and constructions are fixed\footnote{The blocking issue is important especially on higher frequency bands, but for simplicity it is not considered in this paper.}. Hence, they are actually correlated among themselves. 

\begin{figure}[!t]
\centerline{\includegraphics[scale=.25]{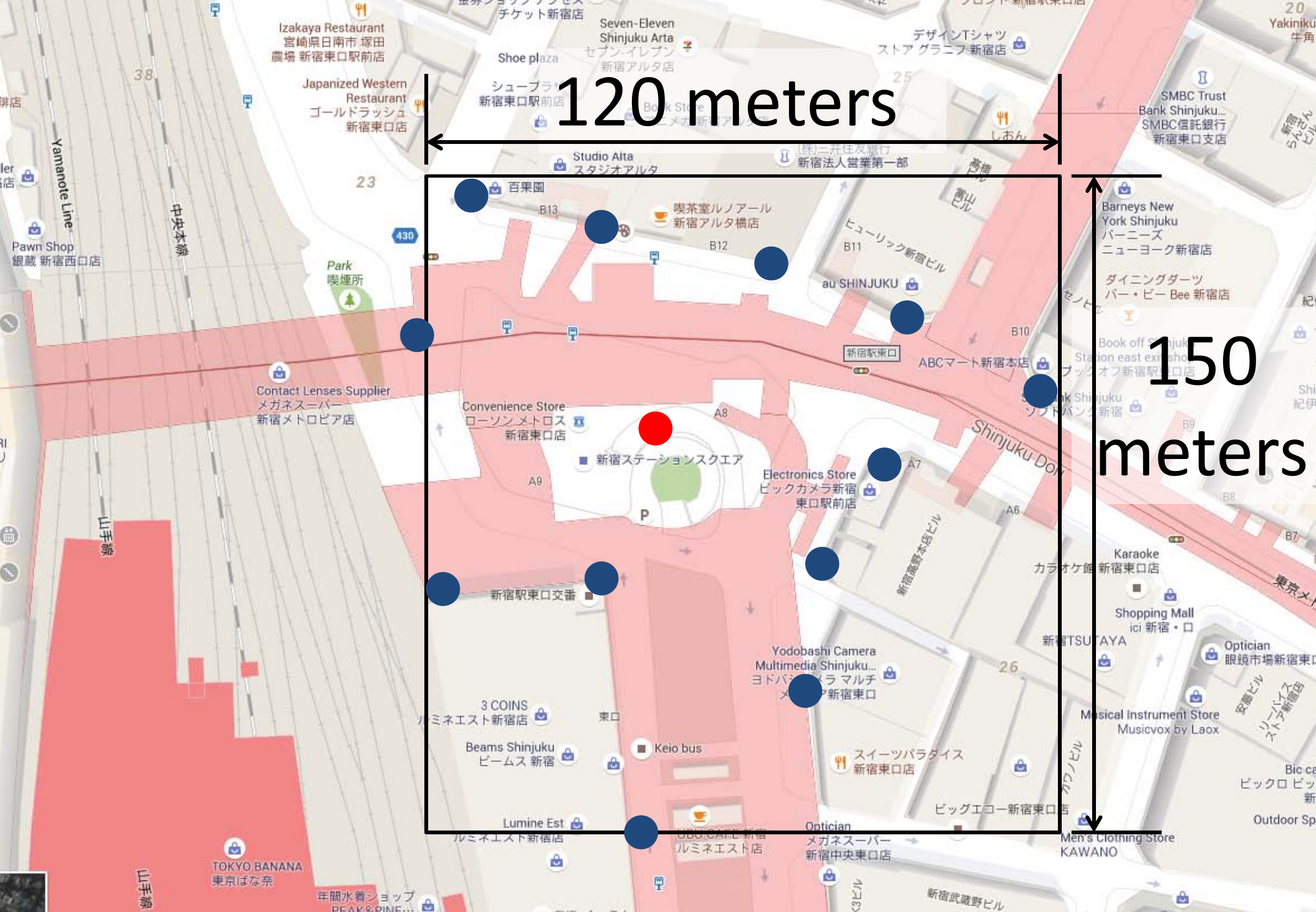}}
\caption{Shinjuku, Tokyo, Japan (The region for study is 150m$\times$120m, bounded by the longitude interval [139.7039,
139.7056] and the latitude interval [35.6880, 35.6894].)}\vspace{-0.15in}
\label{fig:shinjuku}
\end{figure}

In this paper, we explore a machine learning approach to resolve the problem above. In particular, we design and evaluate four distinct neural network models. For each model, we design its corresponding input using the RSS, multipath delays, or their combined use. To enable the machine learning approach, we train and test the models with ray-tracing channel data collected by NTT DOCOMO Research Labs from Shinjuku Square in Tokyo, Japan. Specifically, as shown in Fig. \ref{fig:shinjuku}, we consider a 150m$\times$120m area where there are 13 BSs, 13609 user spots located on a grid, and the distance between any two adjacent user spots on the grid is one meter. For the multipath channel between each BS and each user, the collected measurements on 28GHz carrier frequency include the number of paths,  the RSS and delay of each path,  the (azimuth) AoA and AoD per path (as well as some additional propagation properties  such as the number of reflections and diffractions per path).

\section{Machine Learning Preliminaries}

In this section we introduce several basic concepts for later use, and then describe how to construct the neural network inputs based on the available data introduced in Sec. \ref{sec:problem}.

\subsection{Function Approximation with Neural Networks}

\begin{figure}[!t]
\centerline{\includegraphics[scale=.12]{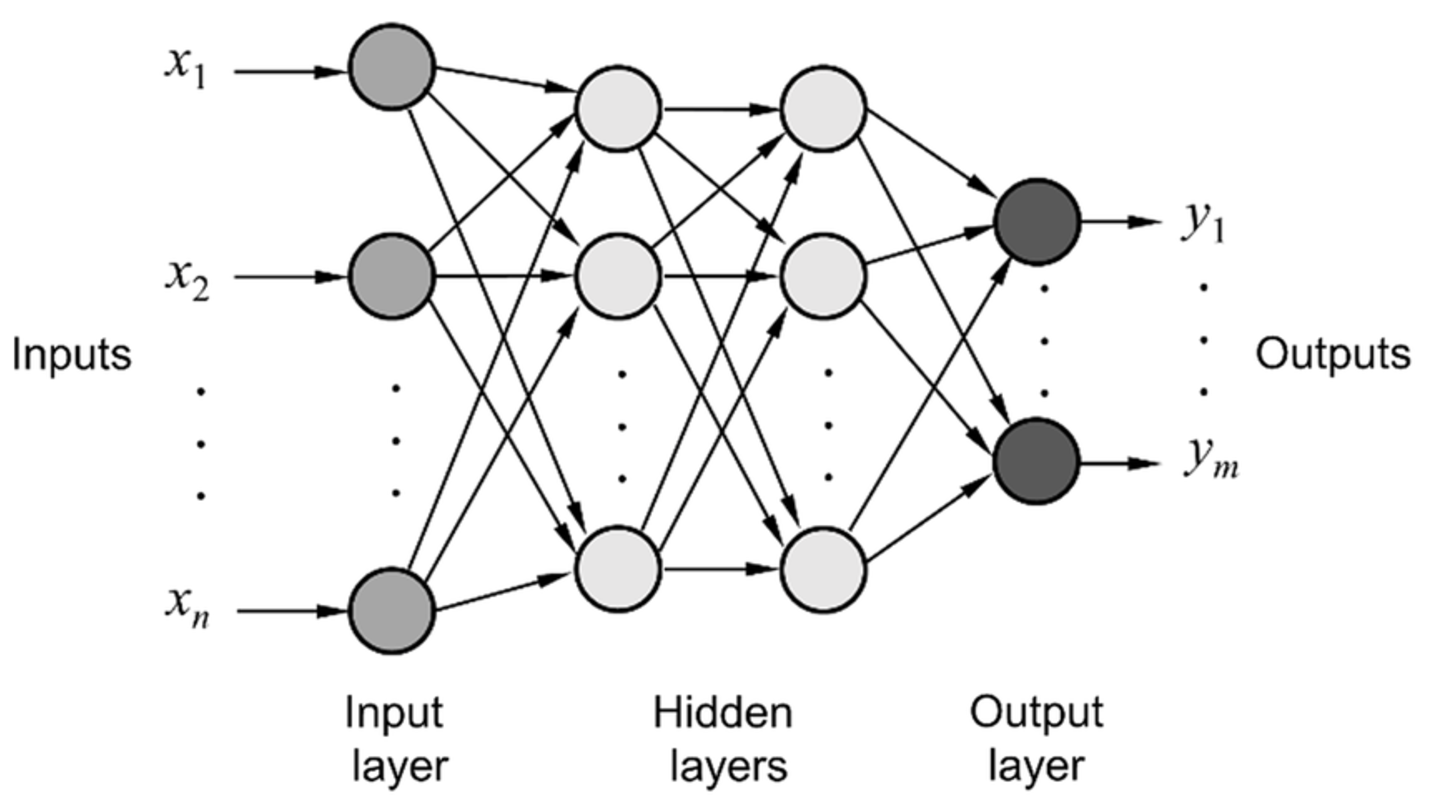}}\vspace{-0.1in}
\caption{A multilayer neural network with two hidden layers.}\vspace{-0.2in}
\label{fig:mlnn}
\end{figure}

The problem above can be recast as function approximation where the goal is to approximate an unknown mapping $f: X\rightarrow Y$ from a set of parameters $X: \{r_l,\tau_l,\theta_l\}_{l=1}^L$ to another set $Y: \{\phi_l\}_{l=1}^L$. In fact, function approximation by employing neural networks has been used extensively in various contexts for learning complicated and non-linear functions such as \cite{cybenko1989approximation}. In particular, as established in \cite{cybenko1989approximation}, \textit{any function} can be approximated up to arbitrary accuracy by a neural network with two hidden layers, as shown in Fig. \ref{fig:mlnn}. Note that each neuron in the hidden and output layers receives an affine transformation (linear combination) of the neurons' values in the preceding layer, and passes it through a nonlinear activation function such as ReLu, sigmoid and $\tanh$ functions. Given enough amount of data, the network can be trained by using the well-known backpropagation algorithm \cite{mitchell1997machine}.

\subsection{Features Representation}\label{sec:IOfeature}

An important aspect of formulating the problem of interest as a supervised learning problem involves constructing appropriate feature representations for the input and output of the neural network.

\begin{figure}[!t]
\centerline{\includegraphics[scale=.4]{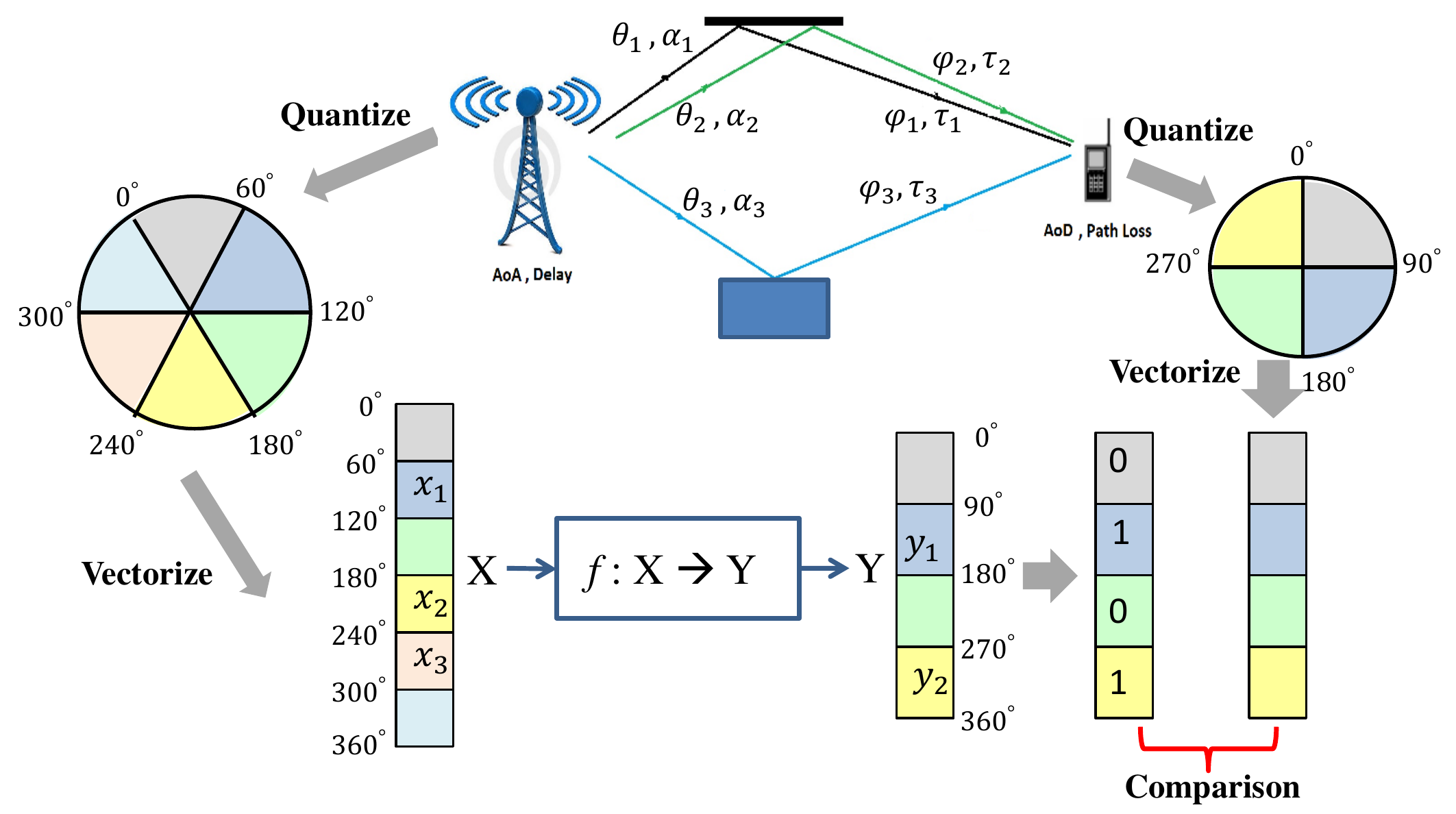}}\vspace{-0.1in}
\caption{Input and output features representation.}\vspace{-0.15in}
\label{fig:rep}
\end{figure}

Let $(\mathbf{x}, \mathbf{y})$ denote the input-output pair of a data sample for training the neural network, where $\mathbf{x}$ is an $N_{\text{BS}}\times 1$ vector, $\mathbf{y}$ is an $N_{\text{MS}}\times 1$ vector, and each entry corresponds to one of their DFT columns. Note that all the entries of $\mathbf{x}$ and $\mathbf{y}$ should be very close to zero, except for those corresponding to the multipath components, i.e., the selected DFT columns indices for beam selection, with significant power. Consider for instance the example in Fig. \ref{fig:rep}.  Suppose that with the angular resolution offered by its $N_{\rm BS}=6$ ULA the BS can distinguish 3 paths, falling into the 2$^{\textrm{nd}}$, 4$^{\textrm{th}}$, and 5$^{\textrm{th}}$ sectors/bins. A $6\times 1$ feature  vector can be created where the corresponding 2$^{\textrm{nd}}$, 4$^{\textrm{th}}$, 5$^{\textrm{th}}$ entries are features of the associated paths, such as delays or RSSs, {\em appropriately renormalized}. In particular, in the case where RSS values are used as features, first the minimum and the maximum RSS values (in dB) in the data set, $r_{\rm min}$ and $r_{\rm max}$, are obtained. For an RSS value of $r$ dB, the entered value in the feature vector is $(r- r_{\rm min})/ (r_{\rm max}- r_{\rm min})$, while zeros are entered for all empty bins. In the case where path delay values are used as features, again the minimum and the maximum path delays in the data set, $\tau_{\rm min}$ and $\tau_{\rm max}$, are obtained.  For a path delay value of $\tau$, the entered value in the feature vector is $(\tau- \tau_{\rm min})/ (2\tau_{\rm max}- \tau_{\rm min})$. In addition, the value 1 (corresponding to a fictitious delay of $2\tau_{\rm max}$ indicating non-existent paths) is entered for all empty bins.

\begin{figure}[!t]
\centerline{\includegraphics[scale=.28]{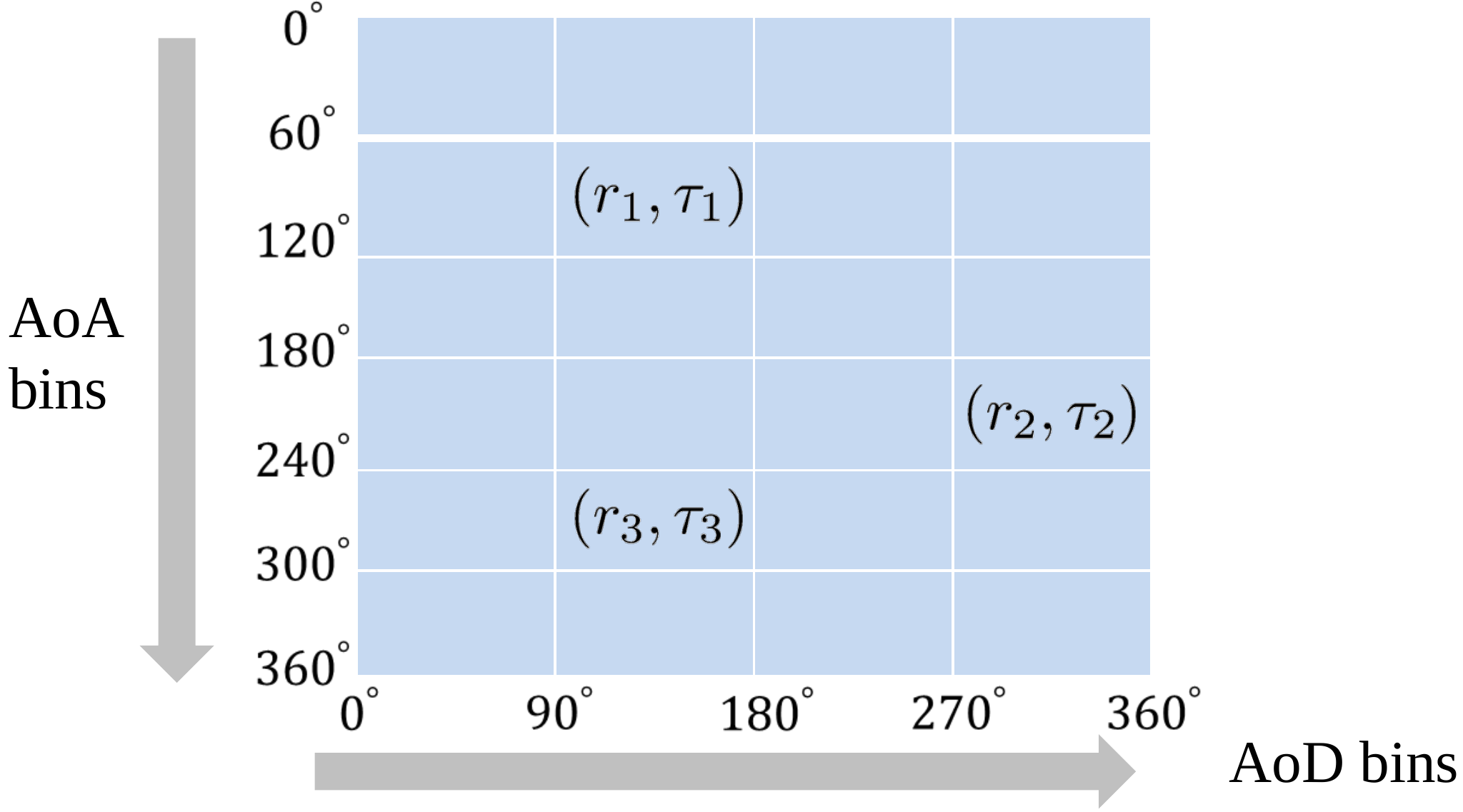}}
\caption{A graphical illustration of the coupling between input and output feature representations.}
\label{fig:io}\vspace{-0.15in}
\end{figure}

At the user, the angular resolution is lower, since it has a smaller array than the BS. In the example depicted in Fig.~\ref{fig:rep}, all the paths fall in two angular bins at the user, and in particular two paths fall into the 2$^{\textrm{nd}}$ bin, and the other path falls into the 4$^{\textrm{th}}$ bin. Thus, we obtain a $4\times 1$ column vector first where the corresponding 2$^{\textrm{nd}}$ and 4$^{\textrm{th}}$ entries are the feature of delay or RSS. Next, we convert this $4\times 1$ vector into another indication vector, where the 2$^{\textrm{nd}}$ and the 4$^{\textrm{th}}$ entries are replaced by 1, indicating the beam index at the user having that AoD component. During the training phase, this information is extracted from the training data set in the form of input-output vector pairs and is used to train the parameters of the neural network. During the testing phase, the trained neural network is fed an input ${\bf x}$ to obtain the estimated output $f({\bf x})$.  We can then evaluate the performance of the trained neural network by comparing the estimate $f({\bf x})$ to the actual output ${\bf y}$ across the test data set using appropriately chosen performance metrics. Fig. \ref{fig:io} illustrates the coupling in the representations of each input-output pair shown in Fig. \ref{fig:rep}. Clearly, a user's features  can be packed into a matrix where the rows (columns) correspond to the AoA (AoD) angular bins of all the associated paths.

\section{Analysis of the Neural Network Model}

In this section, we first conduct correlation analysis on the features representation that we constructed in Sec. \ref{sec:IOfeature} for our ray-tracing data by borrowing the concepts of entropy and mutual information from information theory, so that we can ensure the formulated learning-based inference problem is feasible. Then, we specify a neural network architecture to formulate the estimation problem as an optimization problem. Finally, we  introduce the statistical metrics based on which the neural network performance is evaluated.

\subsection{Correlation Analysis} \label{sec:MI}

Let the random variables $X$ and $Y$ denote the input and the output feature vectors described in Sec. \ref{sec:IOfeature}. Their mutual information is given by $I(Y; X) = H(Y) - H(Y | X)$. To see the mutual information compared to $H(Y)$, we can calculate $I(Y; X) / H(Y)$, which represents the extent of measuring information bits contained in $Y$ given $X$. Clearly, its value falls into the regime [0,1], and the boundaries 0 and 1 correspond to the two extreme cases of independent and fully dependent on one another, respectively. Recall that in our formulations, $X$ is an $N_{\textrm{BS}}\times 1$ vector, and $Y$ is an $N_{\textrm{MS}}\times 1$ vector.

\begin{figure}[!t]
\centerline{\includegraphics[scale=.25]{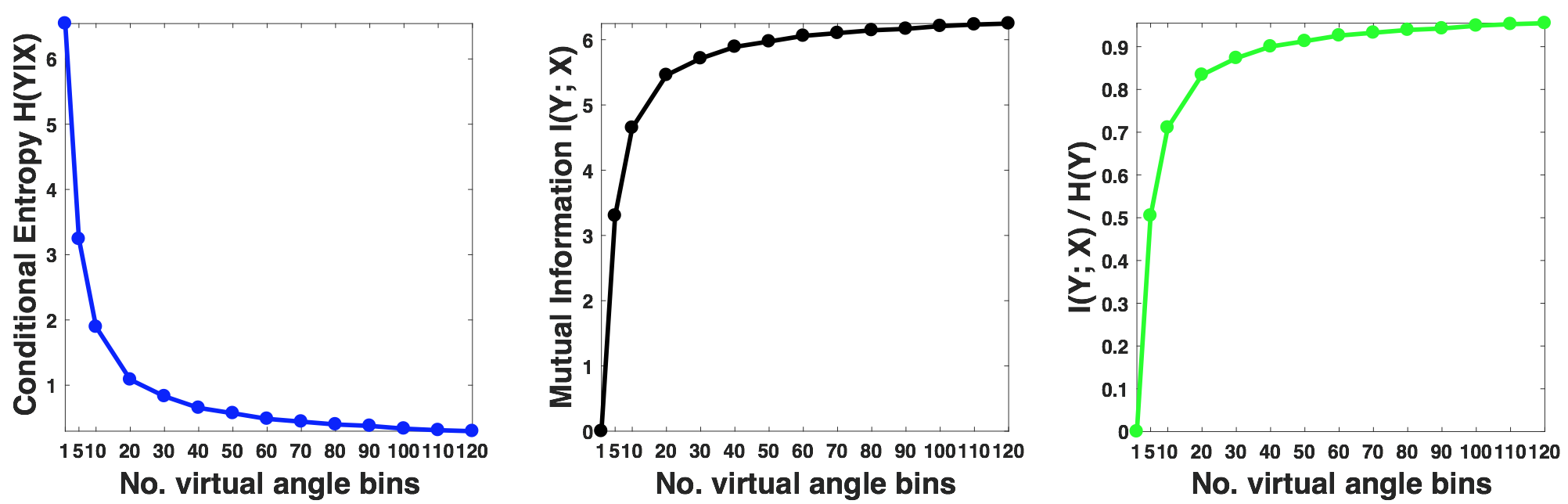}}\vspace{-0.05in}
\caption{Correlation between the input $X$ and the output $Y$: (left) $H(Y | X)$, (middle) $I(Y; X )$, and (right) $I(Y; X )/H(Y)$, as functions of $N_{\textrm{BS}}$ (the dimension of the DFT codebook).}
\label{fig:MI}\vspace{-0.1in}
\end{figure}

Based on our ray-tracing data, in Fig. \ref{fig:MI} we show the values of $I(Y;X)$, $H(Y|X)$ and $I(Y;X) / H(Y)$ as functions of $N_{\textrm{BS}}$, i.e., the number of virtual angle bins at the BS, by fixing $N_{\textrm{MS}}=10$ as an example. Several interesting observations can be made from Fig. \ref{fig:MI}. First, in the leftmost sub-figure, $H(Y | X)$ decreases as $N_{\textrm{BS}}$ increases. Intuitively, this is because when the antenna resolution at the BS is higher, the BS ability to resolve multipaths improves. Next, in the middle one, $I(Y; X)$ grows larger when $N_{\textrm{BS}}$ increases, since $H(Y | X)$ decreases. Finally, in the rightmost one, $I(Y;X) / H(Y)$ increases and is very close to 1 ($\approx 0.97$) when $N_{\rm BS}=120$. Intuitively, it implies that when $N_{\textrm{BS}}$ increases, $Y$ becomes more predictable from $X$ due to the higher antenna resolution.

\subsection{Neural Network Optimization}

Having verified the correlation between the input and the output of the model, we proceed to formulating the estimation problem as an optimization problem and subsequently building a neural network solver. In this paper we restrict our attention to neural networks with only two hidden layers\footnote{Although in principle neural networks with three or more layers are surely worth investigating as solvers for the types of problems we consider in this paper, in the context of the ray tracing data set upon which we based our study  two-hidden layer neural networks performed sufficiently well.}. In addition, the other parameters of the neural network are summarized in Table \ref{table:nn}. Note that the activation function of the output layer is chosen to be linear, because we are solving a regression problem (to estimate the RSS or the delay over AoD bins), and linear transformation is more capable of capturing a wide range of the output values\footnote{If we model the output as a multi-task classification problem where for each task (for each AoD bin), we determine if there exists a path, then the sigmoid function can be chosen as the activation function of the output layer.}.

\begin{table}[!t]
\caption{Neural network architecture.}\vspace{-0.15in}
\begin{center}
\begin{tabular}{|c|c|c|}
\hline
\textbf{Layer}&\textbf{Layer Size}&\textbf{Activation Function} \\
\hline
Input layer & 100 nodes & $-$ \\
\hline
Hidden layer 1 & 50 neurons & Sigmoid \\
\hline
Hidden layer 2 & 20 neurons & Sigmoid \\
\hline
Output layer  & 10 neurons & Linear \\
\hline
\end{tabular}
\label{table:nn}
\end{center}\vspace{-0.15in}
\end{table}
As shown in Table \ref{table:nn}, the neural network consists of the input layer $l=0$, two hidden layers $l=1,2$ and the output layer $l=3$. Let $\bold{W}^{[l,l-1]}$ denote the weight matrix from layer $l-1$ to layer $l$ and $\bold{b}^{l}$ denote the bias at the neurons in layer $l$. In addition, we use $\mathcal{D}_t = \{ (\bold{x}_t^{(n)}, \bold{y}_t^{(n)}) \}_{n=1}^{N_t}$ and $\mathcal{D}_v = \{ (\bold{x}_v^{(n)}, \bold{y}_v^{(n)})\}_{n=1}^{N_v}$ to denote the training and the test sets, respectively, where $N_t = |\mathcal{D}_t|$ and $N_v = |\mathcal{D}_v|$ and $\mathcal{D}_t\cap \mathcal{D}_v=\emptyset$. Moreover, let $h_{W,b}(\bold{x}_t^{(n)})$ denote the output of the neural network in response to the input $\bold{x}_t^{(n)}$. Hence, the cost function during the training phase can be written as

\small\vspace{-0.2in}
\begin{eqnarray}\label{eq:J}
J(\bold{W},\bold{b})\!=\!\frac{1}{N_t}\sum_{n=1}^{N_t}\! \big\| h_{W,b}(\bold{x}_t^{(n)})\! -\! \bold{y}_t^{(n)}\!\big\|^2 \!+\frac{\lambda}{2}\! \sum_{l=1}^{3}\big\|{\bf W}^{[l,l-1]} \big\|_{\textrm{F}}^2,\!\!\!\!
\end{eqnarray}
\normalsize
where $\bold{W}\!=\!\{\bold{W}^{[l,l-1]}\}_{l=1}^3$, $\bold{b}\!=\!\{\bold{b}^{l}\}_{l=1}^3$, $\lambda$ is the regularization factor, and $\|\cdot\|_{\textrm{F}}$ is the Frobenius norm. Training the model involves optimizing the weights $\bold{W}$ and the bias $\bold{b}$ through iterations of the backpropagation algorithm. In this paper, the backpropagation algorithm is implemented using the iterative batch gradient descent optimization algorithm \cite{minFunc}.

\subsection{Statistical Performance Metrics} \label{sec:stat}

To assess the estimation performance of the optimized neural network, we define the following statistical metrics:
\small\vspace{-0.2in}
\begin{eqnarray}
S_t &\!\!\!\!\triangleq\!\!\!\!& \frac{1}{N_t-1} \sum_{n=1}^{N_t} \big\| \bold{y}_t^{(n)} - \overline{\bold{y}}_t \big\|^2,\label{eq:St}\\
\eta_v &\!\!\!\!\triangleq\!\!\!\!& \frac{1}{N_v} \sum_{n=1}^{N_v} \big\| \bold{y}_v^{(n)} - \overline{\bold{y}}_t \big\|^2,\label{eq:etav}\\
\eta_{\textrm{NN}} &\!\!\!\!\triangleq\!\!\!\!& \frac{1}{N_v} \sum_{n=1}^{N_v} \big\| \bold{y}_v^{(n)} - h_{W,b}(\bold{x}_v^{(n)}) \big\|^2.\label{eq:etann}
\end{eqnarray}
\normalsize
Note that $S_t$ in (\ref{eq:St}) denotes the sample variance of the output vectors $\bold{y}_t^{(n)}$ in $\mathcal{D}_t$, where $\overline{\bold{y}}_t\triangleq \frac{1}{N_t} \sum_{n=1}^{N_t} \bold{y}_t^{(n)}$ is the sample average of the output vectors $\bold{y}_t^{(n)}$ in $\mathcal{D}_t$; $\eta_v$ in (\ref{eq:etav}) is the average deviation of the output $\bold{y}_v^{(m)}$ in the test set $\mathcal{D}_v$ from $\overline{\bold{y}}_t$; and $\eta_{\textrm{NN}}$ in (\ref{eq:etann}) represents the empirical approximation of the mean squared  error of the trained neural network estimator when applied to the samples in the test set $\mathcal{D}_v$.

Based on (\ref{eq:St}) and (\ref{eq:etav}), we further define another metric
\begin{eqnarray}
\rho_v \triangleq {\eta_v}/{S_t}.
\end{eqnarray}
It can be seen that $\rho_v$ measures  how well the training data set sample average $\overline{\bold{y}}_t$ estimates the output vectors in the test set $\mathcal{D}_v$ on average, were it to be taken as a coarse estimator of the output vectors in $\mathcal{D}_v$. That is, $\overline{\bold{y}}_t$ can be viewed as a benchmark for performance assessment of any other estimator. Defined as such, $\rho_v>1$ is expected, since the training set sample mean $\overline{\bold{y}}_t$ is not identical to the sample average of the test data set, which minimizes the mean squared error. Moreover, a $\rho_v$ value close to 1 is an indication of the ``closeness'' of the statistical distributions of the output vectors in the training and test sets. Thus, if $\rho_v$ is close to 1, the model that learns the desired mapping based on the samples in $\mathcal{D}_t$ would likely be a good estimator of the outputs in $\mathcal{D}_v$, despite not being exposed to the samples therein during training.

In addition, based on (\ref{eq:etann}) and (\ref{eq:St}), we define the last metric
\begin{align}\label{eq:rhov}
\rho_{\textrm{NN}} \triangleq {\eta_{\textrm{NN}}}/{S_t}.
\end{align}
If the statistical similarity between the data in the training and test sets is sufficiently large (i.e., $\rho_v \rightarrow 1$ ), then $\rho_{\textrm{NN}}$ is expected to be positive and less than 1. The smaller $\rho_{\textrm{NN}}$, the more predictable and reliable the trained estimator is likely to be when applied to the new data samples that are not used to train the model. Thus, the combined use of $\rho_{v}$ and $\rho_{\textrm{NN}}$  helps assess  the estimation performance of our model.

Finally, we also employ the widely used receiver operating characteristic (ROC) curves to statistically evaluate the performance of the optimized neural network based estimator. This is because correct prediction of a dominant AoD bin on the user side can be viewed as {\em correct detection} whereas incorrect identification of an unfavorable AoD bin as a dominant one can be declared as {\em false alarm}. For brevity, we use $P_D$ and $P_F$ to denote their probabilities, respectively.

\section{Simulations} \label{sec:simul}

In this section, we present a statistical performance based evaluation of the neural network techniques we developed. Each neural network was trained and tested using the data collected at BS 8 only (represented by the red dot in Fig. \ref{fig:shinjuku}), which is roughly located at the center of the network and thus can observe a wider angular range. The data set was randomly partitioned into the training set ($10 \%$ of total samples) and the test set ($90 \%$ of total samples)\footnote{Such a partition approach is used for showing the simulation curves only. In the experiments, we partitioned the collected data into $95 \%$ of total samples) for training $5 \%$ for testing to obtain better results.}. Based on the correlation analysis in Sec. \ref{sec:MI}, we choose $N_{\textrm{BS}}=100$ (the size of the DFT codebook at the BS) and $N_{\textrm{MS}}=10$ (the size of the DFT codebook at the MS).  With half-wavelength spaced ULAs at 28GHz, this corresponds to apertures of roughly 53cm and 4.8cm at the BS and the user terminal, respectively.

Given an observed path at the BS, we need to select the most promising bin at the user side from which that path most likely originated.  Recall that the entries in the output vector are either 0 or 1, indicating the angular bins at the MS (see Fig. \ref{fig:rep}). Meanwhile, at the BS, the input vector indicates the (normalized) RSS values or the delays in the corresponding entries.
Using the features representation introduced in Sec. \ref{sec:IOfeature}, we train and evaluate the four distinct neural network schemes, that differ in terms of their inputs:
\begin{enumerate}
\item Each entry of the input vector contains the (normalized) RSS value of the associated angular path. Let $\text{NN}_{r}$ refer to the neural networks trained as such.
\item Each entry of the input vector is given by the delay of the associated angular path. Let $\text{NN}_{\tau}$ refer to the neural networks trained as such.
\item We sequentially train multiple neural networks. Specifically, the outputs of $\text{NN}_{r}$ and $\text{NN}_{\tau}$ are fed to another neural network as inputs. Let us refer to this concatenated neural network architecture as $\text{NN}_{\text{seq}}$.
\item We concatenate the two $100\times 1$ input vectors to $\text{NN}_{r}$ and $\text{NN}_{\tau}$ to form a $200\times 1$ vector, which are then fed into a neural network to estimate the outputs. We denote the neural network trained as such by $\text{NN}_{r,\tau}$.
\end{enumerate}

\begin{table}[!t]
\caption{Statistical assessment of the estimation accuracy of the optimized neural network models.}\vspace{-0.1in}
\begin{center}
\begin{tabular}{|c|c|c|}
\hline
\textbf{Optimized NN}&\textbf{$\rho_v$}&\textbf{$\rho_{\text{NN}}$} \\
\hline
$\text{NN}_{r}$ & 1.0002 & 0.5735 \\
\hline
$\text{NN}_{\tau}$ & 1.0002 & 0.6579 \\
\hline
$\text{NN}_{\text{seq}}$ & 1.0002 & 0.5246 \\
\hline
$\text{NN}_{r,\tau}$ & 1.0002 & 0.5597 \\
\hline
\end{tabular}\vspace{-0.25in}
\label{table:res}
\end{center}
\end{table}

The results of the four schemes above are presented in Table \ref{table:res}. First, it can be seen that $\rho_v = 1.0002$ is promising in the sense that learning the desired mapping from $\mathcal{D}_t$ could potentially lead to accurate estimations. Next, the comparison between the $\rho_{\text{NN}}$ values for $\text{NN}_{r}$ and $\text{NN}_{\tau}$ demonstrates that for the specific ray-tracing data we have, the RSS, at the BS, i.e., the $r_l$'s,  turn out to be the more reliable predictors of the AoD  than the delay, i.e., the $\tau_l$'s.  Our finding differs from what is observed in location estimation problems, where delay based method  outperform RSS-based ones, especially in the presence of large bandwidths. It would be worth evaluating and comparing these networks with real-world measurements, as opposed to ray-tracing data  which are systematically generated by the ray-tracing simulator.

Furthermore, as shown in Table \ref{table:res}, when we combine both the RSSs and the path delays as the input to both $\text{NN}_{\text{seq}}$ and $\text{NN}_{r,\tau}$, the estimation accuracy improves with respect to both $\text{NN}_{r}$ and $\text{NN}_{\tau}$. This is quite intuitive as incorporating more information into the estimation model should, in principle, lead to better estimation performance. Also, note that the structures of the last two neural networks where the RSS and delay features are in the combined use as the input feature are not exactly identical to the first two neural networks in terms of the number of neurons or layers. This motivates us to explore better neural networks with the input features in future.

\begin{figure}[!t]
\centerline{\includegraphics[scale=.38]{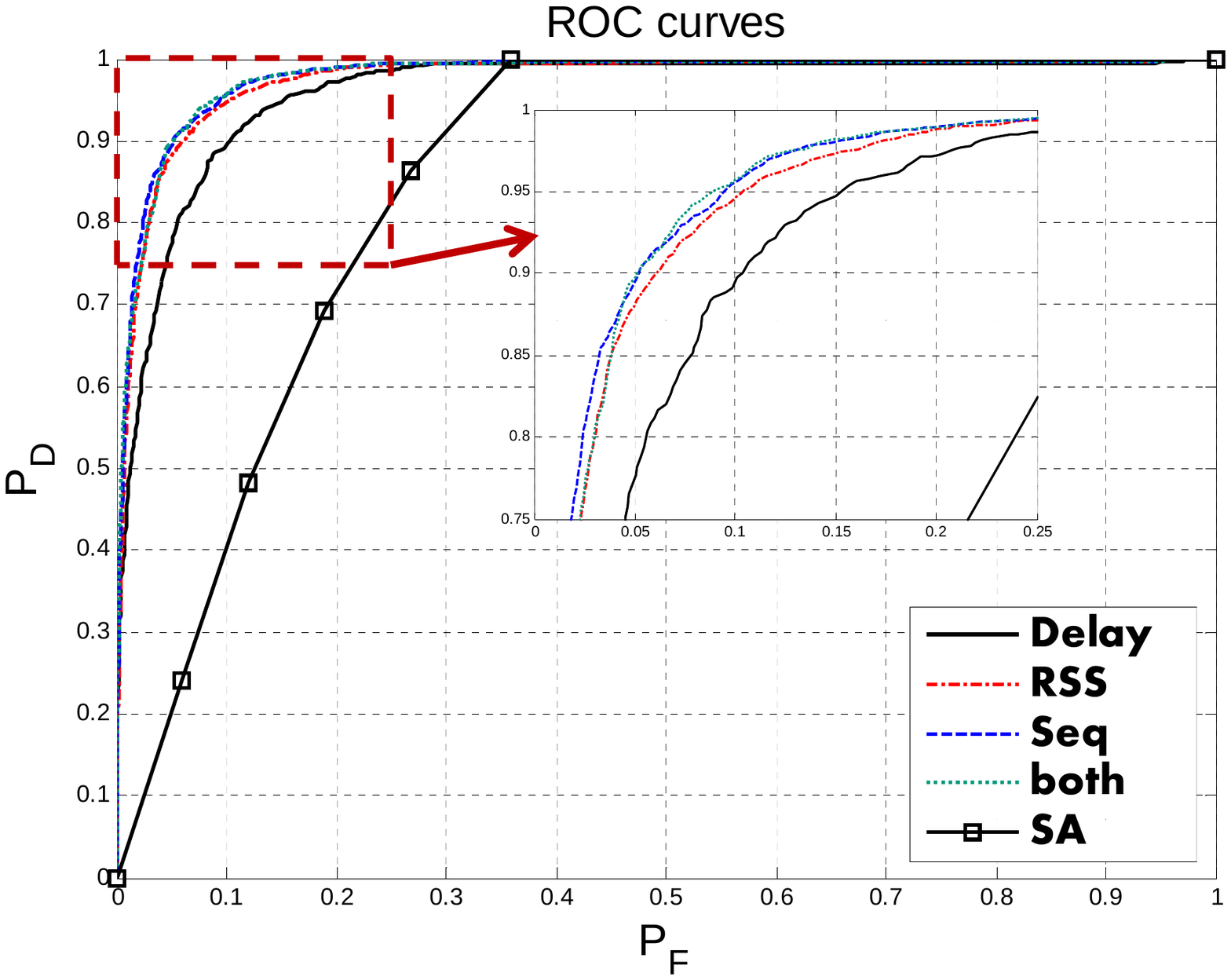}}\vspace{-0.1in}
\caption{ROC curves for various estimation schemes (The horizontal and vertical axes represent the false alarm and detection probabilities, $P_F$ and $P_D$, respectively).}
\label{fig:roc}\vspace{-0.2in}
\end{figure}

Fig. \ref{fig:roc}  shows the ROC  performance of all the schemes under investigation.  In addition, the curve marked ``SA'' in the figure shows the ROC  performance  of the sample average of the output vectors (i.e., $\overline{\bf y}_t$) in the training data set. The sample average can be viewed as a baseline benchmark for the 4 schemes of interest, as it is obtained by using the marginal distribution of the output in case that we do not have any access to the input features.

As shown in Fig. \ref{fig:roc}, the ROC  curves of the other 4 schemes all lie above the sample average curve. Subject to any fixed and small false alarm probability (e.g., $P_F = 0.1$), they all have higher detection probability than the benchmark, which means that they all outperform the sample average (i.e., black-squared ``SA") benchmark estimator. In addition, as shown in the embedded sub-figure, the networks of $\text{NN}_{r}$ (i.e., red-colored ``RSS"), $\text{NN}_{\text{seq}}$ (i.e., blue-colored ``Seq") and $\text{NN}_{r,\tau}$ (i.e., green-colored ``both") are all capable of detecting the target beams with at least $90\%$ probability ($P_D \ge 0.9$) while producing occasional false alarm probability no more than $10\%$ ($P_F \le 0.1$). Finally, the ROC curve of $\text{NN}_{\tau}$ (i.e., black-colored ``Delay") lies beneath the other three curves. This confirms our earlier observation from Table \ref{table:res}  that solely relying on path delay information  leads to less accurate estimation.

\section{Conclusions} \label{sec:recap}

In this paper, we investigated the problem of predicting wireless channel features that are not directly observable at a BS, based on directly observable features and machine learning driven by large amounts of  channel data from the BS's geographical area. In particular, we focused on estimating the dominant virtual angular beams at the user side based on features directly observed at the BS side, such as AoA-specific path RSSs and delays. We developed four learning-based schemes and demonstrated that they can resolve the features of interest with reasonable accuracy by using ray tracing data from Shinjuku Square. Several interesting directions are worthy of further investigation. Training and testing the proposed learning-based estimation schemes with real-world measurement data is necessary to assess their viability in practice. In addition, exploring other machine-learning architectures and then developing beam-selection algorithms are also  goals worth pursuing.

\bibliographystyle{IEEEtran}
\bibliography{REF}

\end{document}